\documentclass{article}
\usepackage{graphics}
\usepackage{amsmath}
\usepackage{amssymb}

\newcommand{\on}{\textrm{on}}
\newcommand{\off}{\textrm{off}}
\newcommand{\eff}{\textrm{eff}}

\newcommand{\open}{\textrm{open}}
\newcommand{\closed}{\textrm{closed}}
\newcommand{\vmedia}{\langle \dot x_{\text{cm}}\rangle}
\newcommand{\sgn}{\text{sgn}}

\title{Information and flux in a feedback controlled Brownian ratchet}

\author{F. J. Cao$^{1,2}$\footnote{Corresponding author: francao@fis.ucm.es}, M.
Feito$^1$ \footnote{feito@fis.ucm.es}, H. Touchette$^3$
\footnote{ht@maths.qmul.ac.uk} \\
$^1$ Departamento de F\'is\'ica At\'omica, Molecular y Nuclear,\\
Universidad Complutense de Madrid, \\ Avenida Complutense s/n,
28040
Madrid, Spain \\
$^2$ LERMA, Observatoire de Paris and CNRS UMR 8112, \\
61, Avenue de l'Observatoire, 75014 Paris, France \\
$^3$ School of Mathematical Sciences, Queen Mary, \\
University of London, London E1 4NS, UK}

\begin{document}
\maketitle

\begin{abstract}
We study a feedback control version of the flashing Brownian
ratchet, in which the application of the flashing potential
depends on the state of the particles to be controlled. Taking the
view that the ratchet acts as a Maxwell's demon, we study the
relationship that exists between the performance of the demon as a
rectifier of random motion and the amount of information gathered
by the demon through measurements. In the context of a simple
measurement model, we derive analytic expressions for the flux
induced by the feedback ratchet when acting on one particle and a
few particles, and compare these results with those obtained with
its open-loop version, which operates without information. Our main finding 
is that the flux in the feedback case has an upper bound proportional
to the square-root of the information. Our results provide a
quantitative analysis of the value of information in feedback
ratchets, as well as an effective description of imperfect or
noisy feedback ratchets that are relevant for experimental applications.
\end{abstract}



\section{Introduction}

Thermal ratchets or Brownian motors can be viewed as controllers
that act on stochastic systems with the aim of inducing directed
motion through the rectification of fluctuations
\cite{mag93,ajd93,ast94,rei02,lin02,cao04,din05}. In most cases,
the system to be controlled is modelled as a collection of
Brownian particles undergoing Langevin dynamics, and the control
action---that is, the rectification mechanism---is implemented by
applying random or deterministic time-dependent perturbations to
the particles. In this context, one can distinguish, as is common
in control theory \cite{sten94}, two types of ratchets: (i)
\emph{open-loop} ratchets, which are ratchets that apply a
rectifying potential independently of the state of the system to
be controlled; (ii) \emph{closed-loop} or \emph{feedback}
ratchets, whose rectification action on a system has an explicit
dependence on that system's evolution in time.

Examples of open-loop ratchets include the flashing ratchet
\cite{ajd93,ast94} and the rocking ratchet \cite{mag93,ast94}. An
example of closed-loop ratchet based on the flashing ratchet was
proposed in \cite{cao04} (see also \cite{din05}). This feedback
flashing ratchet could be implemented experimentally by monitoring
colloidal particles suspended in solution and by exposing the
particles to a saw-tooth dielectric potential as in \cite{rou94},
but with the potential turned on and off depending on particles'
state. The feedback ratchet of \cite{cao04} has also been proposed
as a mechanism to explain the stepping motion of a two-headed
motor protein \cite{bie07}. In a more general context, recent
experiments have shown that information about the location of a
macrocycle in a rotaxane---an organic molecule with a ring
threaded onto an axle---can be used to induce direct transport
away from thermal equilibrium~\cite{ser07}. The operation of such
a molecular ratchet is information-dependent, as it relies on
knowledge of the position of the ring. The use of information is
also relevant in other chemical and biological ratchet-like
systems~\cite{kay07}.

The main motivation for studying closed-loop ratchets is that
they have the potential to perform better as rectifiers of motion
than open-loop ratchets, thereby opening the possibility of improving 
the technological applications of ratchets. Our goal in this paper 
is to establish a quantitative comparison between closed- and open-loop
ratchets that explicitly focuses on what distinguishes them, namely 
the use of information. This is done in three steps using the feedback
ratchet of \cite{cao04} as a case example. First, we show how the
information used by this ratchet can be quantified
(Sec.~\ref{secmodel}). Then we study how the performance of that
ratchet, measured by the magnitude of the flux of particles that
it induces, varies as a function of the amount of information used
in the ratchet effect (Sec.~\ref{secres}). The results obtained
are discussed in Sec.~\ref{secdisc} and compared with those
obtained with the open-loop version of the flashing ratchet, which
operates without information. In Sec.~\ref{secdisc}, we also
discuss the performance of the feedback ratchet as a function of
the correlation established between the controlled system and the
controller, and briefly discuss other performance measures for
feedback ratchets, including the power output and the
thermodynamic efficiency. The results that we obtain are in the
end specific to the feedback ratchet of \cite{cao04}, but the
method that we describe, which is inspired from Maxwell's concept
of thermodynamic demons \cite{lef03}, information theory
\cite{cov91} and the work of one of us \cite{tou00}, is general
and can be applied to other feedback ratchets and other control
systems.

\section{Feedback ratchet and information}
\label{secmodel}

\begin{figure}[t]
\resizebox{4.5in}{!}{\includegraphics{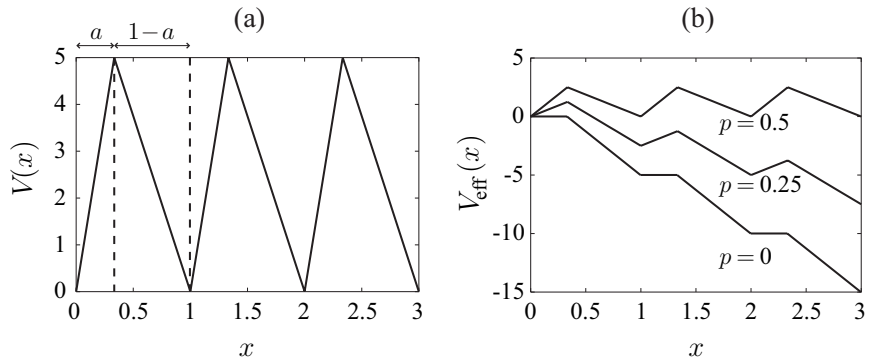}}
\caption{(a) Ratchet potential. (b) Corresponding effective
potential in the case of one particle for three values of the
noise level $p$. The potentials are plotted for $V_0=5$ and
$a=1/3$. Units: $L=1$, $D=1$ and $k_B T=1$.} \label{figpotentials}
\end{figure}

The model of feedback ratchet that we study is constructed as follows \cite{cao04}. We consider a system of $N$ particles with positions $x_i(t)$, whose evolution is described by a set of overdamped Langevin equations:
\begin{equation}
\gamma \dot x_i(t)=\alpha(t)F\left( x_i(t)\right)+\xi_i(t);\quad i=1,\dots,N.
\label{langevin}
\end{equation}
Here, $\gamma$ is the friction coefficient related to the diffusion coefficient $D$ through Einstein's relation $D=k_B T/\gamma$, and $\xi_i(t)$ are Gaussian white noises of zero mean satisfying the fluctuation-dissipation relations $\langle\xi_i(t)\xi_j(t')\rangle =2\gamma k_B T\delta_{ij}\delta(t-t')$. All the particles are subjected to the same potential force $F(x)=-V'(x)$, derived from the following asymmetric, saw-tooth potential:
\begin{equation}
V(x)=
\begin{cases}
\frac{xV_0}{aL} & \mbox{if $0\leq \frac{x}{L}\leq a$} \\
V_0-\frac{V_0}{1-a}\left( \frac{x}{L}-a\right) & \mbox{if $a< \frac{x}{L}\leq 1$},
\end{cases}
\label{potential}
\end{equation}
which is made periodic by the condition $V(x+L)=V(x)$; see Fig.~\ref{figpotentials}(a). Finally, $\alpha(t)$ is a control parameter that switches the potential on ($\alpha=1$) or off ($\alpha=0$). In the open-loop flashing ratchet, the value of $\alpha(t)$ is typically changed periodically in time, whereas in the feedback ratchet of \cite{cao04}, $\alpha(t)$ is set to 1 if the net force
\begin{equation}
f(t)=\frac{1}{N}\sum_{i=1}^N F \left( x_i(t) \right)
\end{equation}
applied to the particles is positive; otherwise, $\alpha(t)=0$. Thus $\alpha(t)=\Theta\left( f(t)\right)$, where $\Theta(x)$ is the Heaviside function. This feedback control strategy is the best possible strategy for maximizing the average velocity of one particle, but not the best strategy when it comes to more than one particle. This can be understood by noting that the system's dynamics tends to get trapped as $N\rightarrow\infty$, with the consequence that the particle flux decreases to zero in this limit \cite{cao04}. By contrast, the open-loop ratchet induces a flux of particles which is independent of the number of particles.

For the remaining, it is useful to picture the ratchet as a
Maxwell's demon \cite{lef03} which rectifies the motion of the
Brownian particles by repeatedly estimating the sign of $f(t)$,
and by subjecting the particles to the on or off potential
depending on the value of the sign measured. When selecting the
potential at a given time $t$, the demon uses only the sign of
$f(t)$ or, equivalently, $\alpha(t)$ at time $t$. It does not use
past information about $f(t)$, nor does it use any detailed
information about the positions of the particles. Accordingly,
what should be quantified as the relevant information used by the
feedback ratchet is the information content or \emph{variability}
of $\alpha(t)$, given by its entropy
\begin{equation}
I=H(b)=-b\log_2 b- (1-b)\log_2 (1-b),
\label{Hbinary}
\end{equation}
where $b$ represents the probability that $f(t)$ is negative. The information $I$ is measured in bits, and represents the \emph{average} information content of $\alpha(t)$ in that it corresponds to the average number of bits needed to store the random outcomes of $\alpha(t)$ \cite{cov91}.

Since our goal is to study the performance of the demon as a
function of $I$, we need to supplement our measure of information
with a way to vary that information. This is accomplished by
introducing noise in the estimation of $f(t)$. It should be said
that noise is always present in control systems in the measurement
step, in the transmission of the measurement information to the
controller, or even in the control-actuation step. Moreover,
adding noise to a ratchet model can provide an effective way of
describing an imperfect feedback controlled ratchet, such as one
plagued by time delays [16].

Here we assume that there is a noise in the estimation of $f(t)$, which leads the demon to wrongly estimate the sign of $f(t)$ with a probability $p\in[0,1/2]$, thereby leading it to apply the wrong potential with probability $p$. Thus, when $f(t) \geq 0$, the demon inadvertently switches off the potential with probability $p$, resulting in an effective on potential $V_{\eff,\on}(x)=(1-p)V(x)$. Conversely, when $f(t)<0$, the demon inadvertently switches on the potential with probability $p$, resulting in an effective off potential $V_{\eff,\off}(x)=pV(x)$. The combination of these two situations leads, in effect, to
having the following ``noisy'' control strategy:
\begin{equation}
\alpha_\eff (t)=(1-p)\Theta(f)+p\Theta(-f).
\label{alpha}
\end{equation}

From the point of view of information theory, the noisy measurement of the sign of $f(t)$ is equivalent to a noisy transmission channel known as the binary symmetric channel \cite{cov91}. The average amount of information transmitted through this channel is measured in terms of a quantity known as the mutual information (see \cite{cov91} for a general definition of this quantity). In our case, the mutual information can be calculated exactly (see Sec.~8.1.4 of \cite{cov91}), and has for expression
\begin{equation}
I=H(q)-H(p),
\label{info1}
\end{equation}
where $H$ is, as in Eq.~\eqref{Hbinary}, the binary entropy function, $p$ is the noise level, and $q$ is the probability that the corrupted sign of $f(t)$ is negative. In terms of the probability $b$ that the \emph{actual} sign of $f(t)$ is negative, we have $q=(1-p)b+p(1-b)$. Note that $b$ depends in general on the number $N$ of particles, the characteristics of the ratchet potential $V(x)$, as well as $p$, so that $I$ is a function of all these parameters.

\section{Results}
\label{secres}

The noise model that we consider is such that the ratchet operates with maximum information when $p=0$, in which case $I=H(b)\leq 1$, and with minimum information ($I=0$) when $p=1/2$. To study the exact performance of the ratchet in and in between these two regimes, we derive in this section the expression of the flux of rectified particles as a function of $p$, and rewrite this expression as a function of $I$ using Eq.~(\ref{info1}). Both the cases of one particle and a few particles are considered.

\subsection{One-particle case}

\begin{figure}[t]
\resizebox{4.5in}{!}{\includegraphics{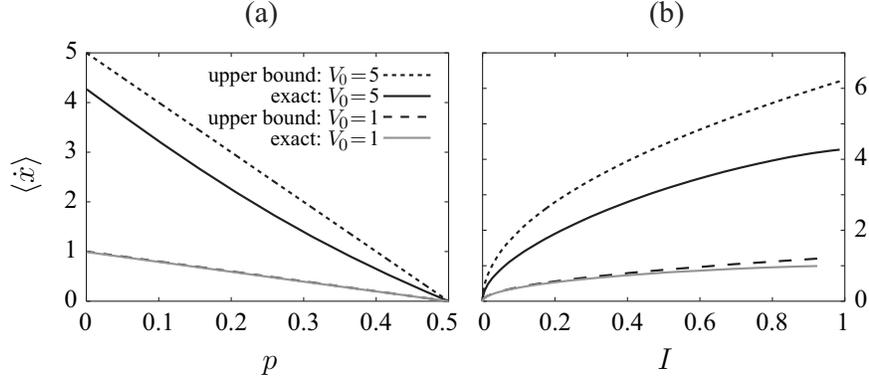}} \caption{(a)
Stationary flux as a function of the noise level $p$ for the
potential heights $V_0=1$ and $V_0=5$ in the one particle case.
(b) Stationary flux as a function of the information $I$. Units:
$L=1$, $D=1$ and $k_B T=1$.} \label{figflux}
\end{figure}

For a single controlled particle, the net force is simply $f(t)=F(x(t))$, with $x(t)$ the position of the particle. Recalling the form of the saw-tooth potential defined in Eq.~\eqref{potential}, we have that $f(t)<0$ for $x\in(0,aL)$, and $f(t)>0$ for $x\in(aL,L)$. As a result, the effective control parameter $\alpha_\eff(t)$ can be
rewritten as
\begin{equation}
\alpha_{\text{eff}}(x) =
\begin{cases}
p       & \mbox{if $0 < \frac{x}{L}\leq a$}\\
1-p     & \mbox{if $a<\frac{x}{L}\leq  1$}.
\end{cases}
\end{equation}
From this expression, we obtain an exact analytical expression for the average flux $\langle \dot x\rangle$ of the particle by solving the Fokker-Planck equation associated with the effective force $F_\eff(x)=\alpha_\eff (x) F(x)$ that derives from the effective control potential depicted in Fig.~\ref{figpotentials}(b). The result in the stationary regime is
\begin{equation}
\langle \dot x \rangle =\frac{p^2(1-p)^2V_0^2A}{ A E - B^+ B^- },
\label{vp}
\end{equation}
with
\begin{eqnarray}
A               & = & 1-e^{(2p-1)V_0} \nonumber\\
B^{\pm}         & = & e^{\pm pV_0}[a(1-p)+p(1-a)]  -e^{\pm(2p-1)V_0}p(1-a)-a(1-p) \nonumber\\
E               & = & a^2(1-p)^2(1-pV_0-e^{-pV_0}) \nonumber\\
                        & & +ap(1-a)(1-p)(1-e^{-pV_0})[1-e^{(1-p)V_0}] \nonumber\\
                        & & +p^2(1-a)^2[1-e^{(1-p)V_0}+(1-p)V_0],
\end{eqnarray}
in units where $L=1$, $D=1$, and $k_B T=1$. We have checked that this result is correct by performing Langevin simulations of the feedback ratchet. As seen in Fig.~\ref{figflux}(a), the flux is maximum for $p=0$ and decreases monotonically to zero as $p$ goes to 1/2.

To transform the expression of $\langle \dot x \rangle$ shown above into a
function of $I$, we need to invert the expression of $I$ shown in
(\ref{info1}) to obtain $p$ as a function of $I$. This requires the expression
of $b$, which is obtained by integrating over the space interval $[0,aL]$ the stationary distribution of the effective Fokker-Planck equation,
\begin{equation}
\begin{split}
b = & \langle \dot x \rangle \left(\frac{a}{pV_0}\right)^2\Bigg\{(1-e^{-pV_0})\Bigg[1+\\
      &\frac{1-e^{pV_0}-e^{pV_0}(1-e^{-V_0(1-p)})\frac{(1-a)p}{(1-p)a}}{e^{-V_0(1-2p)}-1}\Bigg]-pV_0\Bigg\},
\label{b}
\end{split}
\end{equation}
with $\langle \dot x \rangle$ given in (\ref{vp}), and units $L=1$, $D=1$, and $k_B T=1$. This inversion gives the exact result for $\langle\dot x\rangle$ versus $I$, which is plotted in Fig.~\ref{figflux}(b). Unfortunately, we cannot provide a closed-form expression of $\langle\dot x\rangle(I)$ because $p(I)$ seems to have no closed-form expression. However, it is possible to derive a useful approximation of the exact numerical result reported in Fig.~\ref{figflux}(b). Indeed, we can expand $I(p)$ to second order in $p$ around the minimum located at $p=1/2$ to obtain
\begin{equation}
p(I) \approx \frac{1}{2} - \sqrt{\frac{I\ln 2}{8b(1-b)}},
\label{p_of_I_app}
\end{equation}
assuming that $b$ does not depend on $p$. Then, for small potentials, i.e., $V_0 \ll k_B T$, we have $b\approx a$ and
\begin{equation}
\langle \dot x \rangle \approx \frac{V_0}{L \gamma}(1-2p)
\label{smallpot}
\end{equation}
from Eq.~(\ref{vp}), so that
\begin{equation}
\langle \dot x \rangle \approx \frac{V_0}{L \gamma}\sqrt{\frac{I\ln 2}{2 a(1-a)}}.
\label{v_of_I_one}
\end{equation}
This approximation is a priori valid only in the regime where $I \ll 1$ and $V_0 \ll k_B T$, but Fig.~\ref{figflux}(b) shows that it is accurate over the whole range of $I$ even when $V_0 \approx k_B T$. An additional benefit of Eq.~(\ref{v_of_I_one}) is that it is an upper bound on $\langle\dot x \rangle$ versus $I$ for any value of $I$ and any potential height $V_0$. This follows because the right-hand side of (\ref{smallpot}) is an upper bound on the exact result shown in Eq.~(\ref{vp}) \cite{note1}; see Fig.~\ref{figflux}.

\subsection{Few-particle case}

Approximations similar to those given in (\ref{smallpot}) and (\ref{v_of_I_one}) can also be derived for the case where more than one particle is controlled by the feedback ratchet. In the case of a few particles, the net force has a distribution $ \rho(f) $ which can be approximated by a Gaussian distribution,
\begin{equation}
\rho(f) \approx \frac{1}{\sqrt{2\pi\Sigma^2}}\ e^{-\frac{f^2}{2\Sigma^2}} ,
\label{gaussian}
\end{equation}
having a zero mean and variance
\begin{equation}
\Sigma=\frac{V_0}{L\sqrt{a(1-a)N}}.
\end{equation}
This Gaussian approximation is derived under two basic assumptions
[6]: (i) the positions of the particles are statistically
independent; (ii) the probability of finding a particle in a
negative force interval (for example $[0,aL]$) is $a$. It can be
shown that these two assumptions are verified for small potential
even in the presence of noise (i.e., $ p\neq 0 $).

Using the approximation shown in \eqref{gaussian}, we can obtain
an approximate expression for the average center-of-mass velocity
$\vmedia$ as a function of the transmission error $p$ using the
relation
\begin{equation}
\vmedia \approx \frac{1}{\gamma}\int_{-\infty}^\infty \alpha_{\text{eff}}(f) f \rho(f)df,
\end{equation}
with $\alpha_{\text{eff}}(f)$ given by Eq.~\eqref{alpha}. In our case, we find
\begin{equation}
\vmedia \approx \frac{V_0}{L\gamma\sqrt{ 2\pi a(1-a)N }}(1-2p).
\label{flux_small_approx}
\end{equation}
We have compared this results with Langevin simulations and found good fits for small potentials. For $p=0$, the results of \cite{cao04} for $\vmedia$ is recovered. In addition, we have verified numerically that Eq.~(\ref{flux_small_approx}) is an upper bound of the exact flux for any given potential height.

To turn the expression in (\ref{flux_small_approx}) into an expression involving the information $I$, we use the approximation shown in (\ref{p_of_I_app}) again to obtain
\begin{equation}
\vmedia \approx \frac{V_0}{L\gamma\sqrt{2\pi a(1-a) N}}\sqrt{\frac{I\ln 2}{2 b(1-b)}},
\label{v_of_I_few}
\end{equation}
with $b$ the probability of having a negative force in the few particle case. This probability can be approximated as
\begin{equation}
b \approx \sum_{n>aN}^N\binom{N}{n}a^n (1-a)^{N-n}
 \label{b_few}
\end{equation}
using the same assumptions as those involved in the Gaussian
approximation for $ \rho(f) $. The resulting approximation for
$\vmedia$ is similar to the approximation derived for the
one-particle ratchet in that it is a good approximation for small
values of the potential height, in addition to being an upper
bound on the flux for any given potential height. The latter
property was checked numerically. The accuracy of the
approximation for the probability $b$ was also checked
numerically.

\section{Discussion}
\label{secdisc}

(1) The two approximations shown in (\ref{v_of_I_one}) and (\ref{v_of_I_few})
express the performance of the ratchet demon as a function of the information
$I$ that the demon gathers through the noisy measurement of
$\alpha(t)$. Overall, we see from these results that the flux induced by the
demon is maximum when it has maximum information, i.e., $I=H(b)$, and is zero
when it has zero information. This applies both for the one-particle and
few-particle ratchets. In both cases, we further have that the flux decreases
monotonically as $I$ decreases, and that the flux is approximately
proportional to $\sqrt{I}$. The proportionality constant entering in this
relation depends on the system's characteristics, and shows, in the case of a
few particles, an $N$ dependence that has the effect of reducing the flux as
the number of particles is increased. This extra reduction of the flux is
related to the fact that the fluctuations of the force have a smaller amplitude as $N$ grows.

The decrease of flux directly associated with the decrease of information can be explained by noting again that the on and off potentials are partially `mixed' or randomized by the noise. This is particularly evident when $p=1/2$, i.e., when $I=0$. In this case, the demon has a completely random estimate of the sign of $f(t)$ which is uncorrelated with its true sign; hence $I=0$. With the random value of the sign, the demon then applies a random potential to the particles, thereby injecting the noise of the estimation back into the motion of the particles. Such a feedback of estimation noise is often encountered in real control systems, and can be counteracted in various ways. The most common is to rely on past measurements of the controlled system to better estimate its actual state (filtering) \cite{sten94}. For our demon, this would mean acting with memory of past measurements of the sign of $f(t)$, and error-correcting those measurements to avoid inferring the wrong value of the sign of $f(t)$.

(2) The expressions for the center-of-mass velocity as a function of the noise level $p$ or of the information $I$ can be rewritten in terms of the correlation
\begin{equation}
C = \langle \sgn f \ \sgn \tilde f \rangle,
\end{equation}
where $ \sgn f $ is the real sign of the net force that the particles would
feel if the potential were on, while $ \sgn \tilde f $ is the value that the
controller receives. This correlation can be computed by using
\begin{equation}
C = P_{++}+P_{--}-P_{+-}-P_{-+},
\end{equation}
where $P_{ij}$ is the joint probability that the real net force $f$ is positive ($i=+$) or negative ($i=-$) and that the controller receives a positive ($j=+$) or negative ($j=-$) net force $\tilde f$. This joint probability is easily computed knowing that $\sgn f$ is different from $\sgn \tilde f$ with probability $p$. Thus we have $P_{-+}= bp$, $P_{--}=b(1-p)$, $P_{+-}=(1-b)p$ and $P_{++}=(1-b)(1-p)$, where $b$ is the probability of $\sgn f$ being negative, so that
\begin{equation}\label{cor1}
C = 1-2p.
\end{equation}
Using this relation in Eqs.~(\ref{vp}), (\ref{smallpot}) and (\ref{flux_small_approx}), we obtain the expressions for the center-of-mass velocity as a function of the correlation $C$. In particular, for small potentials, we obtain
\begin{equation}
\vmedia \approx \frac{V_0}{L\gamma} \, C
\label{oneC}
\end{equation}
for one particle, and
\begin{equation}
\vmedia \approx \frac{V_0}{L\gamma\sqrt{2\pi a(1-a)N}} \, C
\label{fewC}
\end{equation}
for few particles.

As for Eqs.~\eqref{smallpot} and \eqref{flux_small_approx}, the
expressions shown above are upper bounds for the center-of-mass
velocity for all potential heights, which show that the flux
performance is reduced when the correlation between the controlled
system and the controller decreases. This loss of correlation is
always present in physical systems, and can be due to noise in the
measurement of $f(t)$ or in the transmission of this measurement
to the controller. In this sense, the expressions shown in
\eqref{oneC} and \eqref{fewC} can be thought of as describing a
noisy feedback ratchet for which all the noise sources are
effectively described by $C$. Such an effective description in
terms of $C$ can be used, in addition, to model other imperfect
feedbacks, such as time-delayed feedbacks \cite{fei07b}. In the
latter case, $\tilde f= f (t-\tau)$, where $\tau$ is a positive
time delay, implying a loss of correlation between the actual
force $f=f(t)$ and the applied force $\tilde f$.

Time delays are expected to be present in the experimental
implementation of the feedback ratchet mentioned in the
introduction, in which colloidal particles suspended in a solution
are monitored and exposed to a saw-tooth dielectric potential. The
previous discussion can directly be applied to this situation by
computing $C$ from the time series of $f(t)$ and the delayed
signal $f(t-\tau)$, allowing the use of Eqs.~\eqref{oneC}
and~\eqref{fewC} to predict the effects of time delays on the
flux. The present noisy feedback model could also be useful as an effective description of the operation of an imperfect
feedback loop in other ratchet-like systems \cite{bie07,ser07}.

(3) Experimental implementations of feedback ratchets are
unavoidably imperfect and noisy due to the aforementioned time
delays and other experimental imperfections. 
These real-world limitations can be modelled, to a first level of
approximation, by an effective noise level $p$ acting at the level of estimation. 
With this in mind, one can use our noisy feedback ratchet model as a valuable
effective model for estimating the improvement in flux
that can be obtained in experimental implementations, such as those
proposed in~\cite{cao04,fei07b,cra08}. The authors of
\cite{cra08}, for example, propose a feedback ratchet based on a
scanning line optical trap. From the relevant parameters of their
experimental set-up they found a probability of $1\%$ of
calculating the wrong sign of $f(t)$, and an information content
of about $I=0.9$ bits. Using the results presented here, they
obtain that no more than $95\%$ of the maximum gain achieved
by the feedback strategy can be observed for that real system. The
experimental realization of this system is currently under way
\cite{cra08}.

(4) In general, the flux generated by the open-loop ratchet is much smaller
than the flux generated by the feedback ratchet \cite{cao04}. For the
saw-tooth potential considered here, the \textit{optimal} open-loop
protocol generating the largest flux is the periodic protocol with
on-potential time $T_\on$ and off-potential time $T_\off$. For $V_0=5k_BT$
and $a=1/3$, the optimal values of these times are $T_\on\approx 0.06$ and
$T_\off\approx 0.05$, in units where $D=1$ and $L=1$, yielding
$\vmedia_\open\approx 0.3$. For the one-particle case, we have by
comparison $\vmedia_\closed\approx 4.3$ when $I$ is maximum. For other
values of $I$, the previous results for one and for few particles state
that the center-of-mass flux $\vmedia$ is upper-bounded by $ M \sqrt{I}$,
where $ M $ is a constant depending on the system's characteristics; this
upper-bound is also greater than the open-loop value for most values of
$I$. Therefore, we can write

\begin{equation}
\vmedia_\closed-\vmedia_\open\leq M \sqrt{I}.
 \label{ineq1}
\end{equation}

The feedback protocol that we have considered, which performs an
instantaneous maximization of the center-of-mass velocity
\cite{cao04}, is the optimal protocol that maximizes the flux in the
one particle case for a noiseless channel ($p=0$). We expect this
protocol to give a flux close to the maximum possible value in the
few particles case and in the presence of noise with a memoryless
protocol (note that protocols with memory can perform error
correction). Therefore, we expect Eqs.~\eqref{v_of_I_one} and
\eqref{v_of_I_few} to be upper bounds of the maximum flux that can
be obtained with a memoryless closed-loop control protocol that
uses an amount of information $ I $ about the system. Similarly,
the inequality shown in Eq.~\eqref{ineq1} is expected to set an
upper-bound on the maximum improvement that can result from
changing an open-loop protocol to a memoryless closed-loop
protocol. This, at least, is the case for one particle, as the
instant maximization protocol is optimal when applied to one
particle. For the few particle case, we expect the inequality to
hold, although it could be violated by protocols other than the
one considered here, as these could potentially be more efficient.

(5) We have focused here on proving an upper bound for the particle flux
because this quantity can readily be measured in experimental realizations
of feedback ratchets \cite{cra08}. In a recent paper~\cite{fei07}, written after the
present one, an analogous upper bound was derived for the power output of a 
feedback ratchet, based on the results and techniques presented here. 
The difference between the particle flux and the 
power output is that the latter quantity requires that we impose
a constant load force against the flux so as to compute the work 
done against the load; see \cite{fei07} for more details.

(6) Another important performance measure for ratchets
is the efficiency~\cite{par98}. For the computation of this quantity, it is 
important to note that feedback ratchets have an extra energy input 
compared to open-loop ratchets, related to the fact that information
has an energy cost \cite{lef03}.
This energy cost, known as Landauer's erasure cost because it is incurred when
information is erased, effectively prevents Maxwell's demon-type engines,
like feedback ratchets, to have efficiencies greater than one (as required by
the second law of thermodynamics) \cite{lef03}. 
The calculation of this energy cost requires the computation of 
the mutual information between the controlled system and the controller, 
conditioned on the past history of the controller's evolution~\cite{tou00,cao08}. The role of the 
conditioning is to take into account the correlations between the measurements, 
and to avoid redundancies in the computation of the entropy reduction. The conditioning
is also consistent with the fact that the controller's measurement record, seen as blocks
of bits, must be compressed before it is erased in order to minimize the erasure 
cost; see Ref.~\cite{zur89}.

(7) We have not addressed the many particle case, i.e., the case
where the fluctuations in the net force are smaller than the
typical values of the net force. The reason for this omission is
that, in the many particle case, the maximum increase of
performance that results from changing the optimal open-loop
protocol to a closed-loop protocol is negligible.

\section{Summary}
\label{secsum}

In summary, we have quantified the information gathered by a
feedback control ratchet, and have derived analytical upper
bounds, expressed as a function of the information, which
establish limits on the difference between the flux of particles
created by a closed-loop, flashing ratchet and the flux created by
its open-loop version. These bounds provide a direct evaluation of
the performance of the feedback ratchet as a function of the
information that it uses, and make more precise the idea that
feedback ratchets act like Maxwell's demons that use information
about the state of particles to rectify the particles' motion. In
addition, the analytic results found for the flux are useful in
that they provide an effective description of a feedback flashing
ratchet affected either by noise in the measurement process or
other imperfections in the feedback mechanism. This effective
description is useful for predicting the results of experimental
realizations of feedback ratchets, as any experimental realization
is subjected to noises, delays, and other imperfections in the
feedback.

\section*{Acknowledgments}
This work was supported by MCYT (Spain)
(Grants BFM2003-02547/FISI, FIS2005-24376-E, and FIS/2006-05895),
and by the ESF program STOCHDYN. M.F.\ acknowledges support from
UCM (Spain) through the grant Beca Complutense. H.T.\ was
supported by NSERC (Canada), the Royal Society, and HEFCE
(England).


\begin{thebibliography}{99}

\bibitem{mag93} M. O. Magnasco, Phys. Rev. Lett. \textbf{71}, 1477 (1993).

\bibitem{ajd93} A. Ajdari and J. Prost, C. R. Acad. Sci. Paris II \textbf{315}, 1635 (1993).

\bibitem{ast94} R.D. Astumian and M. Bier, Phys. Rev. Lett. \textbf{72}, 1766 (1994).

\bibitem{rei02} P. Reimann, Phys. Rep. \textbf{361}, 57 (2002).

\bibitem{lin02} H. Linke, Appl. Phys. A \textbf{75}, 167 (2002).

\bibitem{cao04} F. J. Cao, L. Dinis, and J. M. R. Parrondo, Phys. Rev. Lett. \textbf{93}, 040603 (2004).

\bibitem{din05} L. Dinis, J. M. R. Parrondo, and F. J. Cao, Europhys. Lett. \textbf{71}, 536 (2005); M. Feito and F. J. Cao, Phys. Rev. E \textbf{74}, 041109 (2006).

\bibitem{sten94} R. F. Stengel, \textit{Optimal Control and Estimation} (Dover, New York, 1994). See also J. Bechhoefer, Rev. Mod. Phys. \textbf{77}, 783 (2005).

\bibitem{rou94} J. Rousselet, L. Salome, A. Ajdari, and J. Prost, Nature {\bf 370}, 446 (1994).

\bibitem{bie07} M. Bier, Biosystems {\bf 88}, 301 (2007).

\bibitem{ser07} V. Serreli, C.-F. Lee, E. R. Ray, and D. Leigh, Nature (London) \textbf{445}, 523 (2007).

\bibitem{kay07} E. A. Kay, D. A. Leigh, and F. Zerbetto, Angew. Chem. Int. Ed. \textbf{46}, 72 (2007).

\bibitem{lef03} H. S. Leff and A. F. Rex, \textit{Maxwell's Demon: Entropy, Classical and Quantum Information, Computing} (Institute of Physics, Bristol, 2003).

\bibitem{cov91} T. M. Cover and J. A. Thomas, \emph{Elements of Information Theory} (John Wiley, New York, 1991).

\bibitem{tou00} H. Touchette and S. Lloyd, Phys. Rev. Lett. \textbf{84}, 1156 (2000); Physica A \textbf{331}, 140 (2004).

\bibitem{fei07b} M. Feito and F. J. Cao, Phys. Rev. E {\bf 76}, 061113 (2007).

E. M. Craig, B. R. Long, J. M. R. Parrondo, and H. Linke,
Europhys. Lett. {\bf 81}, 10002 (2008).

M. Feito and F. J. Cao, Physica A \textbf{387}, 4553 (2008).

\bibitem{cra08} E. M. Craig, N. J. Kuwada, B. J. Lopez, and H.
Linke, Ann. Phys. {\bf 17}, 115 (2008).

\bibitem{note1} The upper bound on $\langle \dot x \rangle$, taken as a function of $p$, was verified numerically.

\bibitem{fei07} M. Feito and F. J. Cao, Eur. Phys. J. B {\bf 59}, 63 (2007).

\bibitem{par98} J. M. R. Parrondo, J.M. Blanco, F. J.  Cao, R. Brito, Europhys. Lett. \textbf{43}, 248 (1998).

\bibitem{cao08} F. J. Cao, M. Feito, \emph{Thermodynamics of feedback 
controlled systems}, arXiv:0805.4824 (2008).

\bibitem{zur89} W. H. Zurek, Phys. Rev. A {\bf 40}, 4731 (1989); Nature {\bf
    341}, 119 (1989).

\end{thebibliography}
\end{document}